\documentclass[preprint,12pt]{elsarticle}

\usepackage{dcolumn}

\usepackage{graphicx}

\biboptions{sort&compress}

\begin{document}

\begin{frontmatter}

\title{On the symmetry of three identical interacting particles in a
one-dimensional box}

\author{Paolo Amore\dag} \ead{paolo.amore@gmail.com}

\author{Francisco M.
Fern\'andez\ddag}\ead{fernande@quimica.unlp.edu.ar}

\address{\dag\ Facultad de Ciencias, CUICBAS,
Universidad de Colima, Bernal D\'{\i}az del Castillo 340, Colima,
Colima, Mexico}

\address{\ddag\ INIFTA (UNLP, CCT La Plata-CONICET), Divisi\'{o}n Qu\'{i}mica Te\'{o}rica,
Diag. 113 y 64 (S/N), Sucursal 4, Casilla de Correo 16, 1900 La
Plata, Argentina}

\begin{abstract}
We study a quantum-mechanical system of three particles in a one-dimensional
box with two-particle harmonic interactions. The symmetry of the system is
described by the point group $D_{3d}$. Group theory greatly facilitates the
application of perturbation theory and the Rayleigh-Ritz variational method.
A great advantage is that every irreducible representation can be treated
separately. Group theory enables us to predict the connection between the
states for the small box length and large box length regimes of the system.
We discuss the crossings and avoided crossings of the energy levels as well
as other interesting features of the spectrum of the system.
\end{abstract}


\end{frontmatter}

\section{Introduction}

\label{sec:intro}

During the last decades there has been great interest in the model of a
harmonic oscillator confined to boxes of different shapes and sizes\cite
{AK40,A41,A42,C43,AK45,D52,BS55,D66,V68,V73,CF76,AM77,R78,ALZ80,AILZ81,
BR81,FC81a,FC81b,AGZL83,CM83,ML83,A97,V98}. Such model has been suitable for
the study of several physical problems ranging from dynamical friction in
star clusters\cite{C43} to magnetic properties of solids\cite{D52} and
impurities in quantum dots\cite{V98}.

One of the most widely studied models is given by a particle confined to a
box with impenetrable walls at $-L/2$ and $L/2$ bound by a linear force that
produces a parabolic potential-energy function $V(x)=k(x-x_{0})^{2}/2$,
where $|x_{0}|<L/2$. When $x_{0}=0$ the problem is symmetric and the
eigenfunctions are either even or odd; such symmetry is broken when $%
x_{0}\neq 0$. Although interesting in itself, this model is rather
artificial since the cause of the force is not specified. It may, for
example, arise from an infinitely heavy particle clamped at $x_{0}$. For
this reason we have recently studied the somewhat more interesting and
realistic case in which the other particle also moves within the box\cite
{AF10}. Such a problem is conveniently discussed in terms of its symmetry
point-group; for example: it is $C_{i}$ when the two particles are different
and $C_{2h}$ for identical ones.

It follows from what was just said that the case of identical particles is
of greater interest from the point of view of symmetry. For this reason in
this paper we analyse the model of three interacting particles confined to a
one--dimensional box with impenetrable walls. In Section~\ref
{sec:general_trap} we consider three particles in a general one-dimensional
trap with two-particle interactions and discuss its symmetry as well as
suitable coordinates for its treatment. In Section~\ref{sec:1D_Box} we focus
on the case that the trap is given by a box with impenetrable walls and
apply perturbation theory based on the exact results for infinitely small
box length. In Section~\ref{sec:infinite_box} we discuss the limit of
infinite box length as well as the connection between both regimes. In
Section~\ref{sec:RR} we construct a symmetry-adapted basis set for the
application of the Rayleigh-Ritz variational method and discuss the main
features of the spectrum. Finally, in Section~\ref{sec:conclusions} we
provide further comments on the main results of the paper and draw
conclusions.

\section{Three particles in a one-dimensional trap}

\label{sec:general_trap}

We consider three structureless particles in a one-dimensional trap with a
Hamiltonian of the form
\begin{eqnarray}
H &=&-\frac{\hbar ^{2}}{2m}\left( \frac{\partial ^{2}}{\partial x_{1}^{2}}+%
\frac{\partial ^{2}}{\partial x_{2}^{2}}+\frac{\partial ^{2}}{\partial
x_{3}^{2}}\right) +V(x_{1})+V(x_{2})+V(x_{3})  \nonumber \\
&&+W(|x_{1}-x_{2}|)+W(|x_{2}-x_{3}|)+W(|x_{3}-x_{1}|),  \label{eq:H}
\end{eqnarray}
where $V(x_{i})$ confines each particle in a given space region (trap) and $%
W(|x_{i}-x_{j}|)$ are two-body interactions that couple the particles. It is
convenient to define dimensionless coordinates $%
(x,y,z)=(x_{1}/L,x_{2}/L,x_{3}/L)$, where $L$ is a suitable length unit. The
resulting dimensionless Hamiltonian is
\begin{eqnarray}
H^{\prime } &=&\frac{2mL^{2}}{\hbar ^{2}}H=-\left( \frac{\partial ^{2}}{%
\partial x^{2}}+\frac{\partial ^{2}}{\partial y^{2}}+\frac{\partial ^{2}}{%
\partial z^{2}}\right) +v(x)+v(y)+v(z)  \nonumber \\
&&+\lambda \left[ w(|x-y|)+w(|y-z|)+w(|z-x|)\right] ,
\label{eq:H_dimensionless}
\end{eqnarray}
where $v(x)=2mL^{2}V(Lx)/\hbar ^{2}$ and $\lambda
w(|x-y|)=2mL^{2}W(L|x_{1}-x_{2}|)/\hbar ^{2}$, etc. In this equation $%
\lambda $ is a dimensionless parameter that measures the strength of the
coupling interaction.

From now on we omit the prime and simply write $H$ instead of $H^{\prime }$.
In addition to it, we assume that the trap is symmetric: $v(-q)=v(q)$. The
Hamiltonian $H_{0}=H(\lambda =0)$ is invariant under the following
transformations
\begin{eqnarray}
(x,y,z) &\rightarrow &\{x,y,z\}_{P}  \nonumber \\
(x,y,z) &\rightarrow &\{-x,y,z\}_{P}  \nonumber \\
(x,y,z) &\rightarrow &\{x,-y,z\}_{P}  \nonumber \\
(x,y,z) &\rightarrow &\{x,y,-z\}_{P}  \nonumber \\
(x,y,z) &\rightarrow &\{-x,-y,z\}_{P}  \nonumber \\
(x,y,z) &\rightarrow &\{-x,y,-z\}_{P}  \nonumber \\
(x,y,z) &\rightarrow &\{x,-y,-z\}_{P}  \nonumber \\
(x,y,z) &\rightarrow &\{-x,-y,-z\}_{P},  \label{eq:Trans_Oh}
\end{eqnarray}
where $\{a,b,c\}_{P}$ denotes the different permutations of the three real
numbers $a$, $b$ and $c$\cite{F13b}. Note that the $48$ coordinate
transformations (\ref{eq:Trans_Oh}) form a group that is commonly named $%
O_{h}$\cite{H62,C90}. In the particular case that $v(q)=q^{2}$ the symmetry
of $H_{0}$ is given by the full rotation group $O(3)$\cite{H62}.

When $\lambda \neq 0$ the Hamiltonian is invariant under the $12$ coordinate
transformations
\begin{eqnarray}
(x,y,z) &\rightarrow &\{x,y,z\}_{P}  \nonumber \\
(x,y,z) &\rightarrow &\{-x,-y,-z\}_{P},  \label{eq:Trans_D3d}
\end{eqnarray}
and we can describe the symmetry of the system in configuration space by
means of the point group $D_{3d}$\cite{H62,C90}. However, if $v(q)=q^{2}$
the symmetry increases because the trapping potential is invariant under
arbitrary rotations about any axis in configuration space. The coordinate
transformation $(x,y,z)=(q_{1},q_{2},q_{3})\cdot \mathbf{J}$, where\cite{H12}
\begin{equation}
\mathbf{J}=\left(
\begin{array}{lll}
\frac{1}{\sqrt{2}} & -\frac{1}{\sqrt{2}} & 0 \\
\frac{1}{\sqrt{6}} & \frac{1}{\sqrt{6}} & -\frac{2}{\sqrt{6}} \\
\frac{1}{\sqrt{3}} & \frac{1}{\sqrt{3}} & \frac{1}{\sqrt{3}}
\end{array}
\right) ,  \label{eq:J}
\end{equation}
makes the coupling term independent of $q_{3}$ because
\begin{equation}
(x-y,y-z,z-x)=\left( \sqrt{2}q_{1},\frac{\sqrt{6}}{2}q_{2}-\frac{\sqrt{2}}{2}%
q_{1},-\frac{\sqrt{2}}{2}q_{1}-\frac{\sqrt{6}}{2}q_{2}\right) .
\label{eq:Trans_x-y}
\end{equation}
Since the coupling term is therefore invariant under rotations about the $%
q_{3}$ axis by angles of $\frac{2\pi j}{6}$, $j=1,2,3,4,5$ the resulting
Hamiltonian operator
\begin{eqnarray}
H &=&-\left( \frac{\partial ^{2}}{\partial q_{1}{}^{2}}+\frac{\partial ^{2}}{%
\partial q_{2}{}^{2}}+\frac{\partial ^{2}}{\partial q_{3}{}^{2}}\right)
+q_{1}{}^{2}+q_{2}{}^{2}+q_{3}{}^{2}  \nonumber \\
&&+\lambda \left[ w\left( \sqrt{2}\left| q_{1}\right| \right) +w\left(
\left| \frac{\sqrt{6}}{2}q_{2}-\frac{\sqrt{2}}{2}q_{1}\right| \right)
+w\left( \left| \frac{\sqrt{2}}{2}q_{1}+\frac{\sqrt{6}}{2}q_{2}\right|
\right) \right] ,  \nonumber \\
&&  \label{eq:H_dim_D6h}
\end{eqnarray}
is separable into its $(q_{1},q_{2})$ and $q_{3}$ parts and exhibits
symmetry $D_{6h}$.

If $w(|s-t|)=(s-t)^{2}$ the resulting Hamiltonian
\begin{equation}
H=-\left( \frac{\partial ^{2}}{\partial q_{1}{}^{2}}+\frac{\partial ^{2}}{%
\partial q_{2}{}^{2}}+\frac{\partial ^{2}}{\partial q_{3}{}^{2}}\right)
+q_{1}{}^{2}+q_{2}{}^{2}+q_{3}{}^{2}+3\lambda \left(
q_{1}^{2}+q_{2}^{2}\right)  \label{eq:H_dim_D_inf_h}
\end{equation}
is fully separable and exhibits symmetry $D_{\infty h}$\cite{C90}.

\section{One-dimensional box}

\label{sec:1D_Box}

As stated in the introduction we are interested in a system of three
identical particles in a one-dimensional box. To this end we resort to the
Hamiltonian (\ref{eq:H_dimensionless}) and choose the trapping potential
\begin{equation}
v(q)=\left\{
\begin{array}{c}
0\mathrm{\ if\ }|q|<1 \\
\infty \mathrm{\ elsewhere}
\end{array}
\right. .  \label{eq:v(q)_box}
\end{equation}
Therefore, the boundary conditions for the solutions to the Schr\"{o}dinger
equation $H\psi =E\psi $ become
\begin{equation}
\psi (\pm 1,y,z)=\psi (x,\pm 1,z)=\psi (x,y,\pm 1)=0.  \label{eq:box_bc}
\end{equation}
For concreteness we focus the discussion on the Hamiltonian
\begin{equation}
H=-\left( \frac{\partial ^{2}}{\partial x^{2}}+\frac{\partial ^{2}}{\partial
y^{2}}+\frac{\partial ^{2}}{\partial z^{2}}\right) +\lambda \left[
(x-y)^{2}+(y-z)^{2}+(z-x)^{2}\right] ,  \label{eq:H_model}
\end{equation}
that is an extension of the two-particle model discussed earlier\cite{AF10}.

This problem is exactly solvable when $\lambda =0$ and the eigenvalues and
eigenfunctions of $H_{0}$ are
\begin{eqnarray}
E_{n_{1}n_{2}n_{3}} &=&\frac{\pi ^{2}}{4}(n_{1}^{2}+n_{2}^{2}+n_{3}^{2}),%
\;n_{1},n_{2},n_{3}=1,2,\ldots  \nonumber \\
\varphi _{n_{1}n_{2}n_{3}}(x,y,z) &=&\phi _{n_{1}}(x)\phi _{n_{2}}(y)\phi
_{n_{3}}(z)  \nonumber \\
\phi _{n}(q) &=&\sin \frac{n\pi (q+1)}{2}.  \label{eq:H0_eigenval_eigenfun}
\end{eqnarray}
All the sets of quantum numbers produced by the distinct permutations $%
\{n_{1},n_{2},n_{3}\}_{P}$ lead to the same energy (degenerate eigenstates).
In addition to it, there is a Pythagorean degeneracy\cite{F13b} that we do
not discuss here. Each of the functions $\phi _{n}(q)$ is even ($e$) or odd (%
$o$) when $n$ is either odd or even, respectively. A straightforward
analysis based on group theory shows that\cite{F13b}
\begin{eqnarray}
\{e,e,e\}_{P} &\rightarrow &A_{1g}  \nonumber \\
\{e^{\prime },e,e\}_{P} &\rightarrow &A_{1g}\oplus E_{g}  \nonumber \\
\{e^{\prime },e^{\prime \prime },e^{\prime \prime \prime }\}_{P}
&\rightarrow &A_{1g}\oplus A_{2g}\oplus E_{g}\oplus E_{g}  \nonumber \\
\{o,e,e\}_{P} &\rightarrow &A_{2u}\oplus E_{u}  \nonumber \\
\{o,e^{\prime },e^{\prime \prime }\}_{P} &\rightarrow &A_{1u}\oplus
A_{2u}\oplus E_{u}\oplus E_{u}  \nonumber \\
\{e,o,o\}_{P} &\rightarrow &A_{1g}\oplus E_{g}  \nonumber \\
\{o,o^{\prime },e\}_{P} &\rightarrow &A_{1g}\oplus A_{2g}\oplus E_{g}\oplus
E_{g}  \nonumber \\
\{o,o,o\}_{P} &\rightarrow &A_{2u}  \nonumber \\
\{o^{\prime },o,o\}_{P} &\rightarrow &A_{2u}\oplus E_{u}  \nonumber \\
\{o,o^{\prime },o^{\prime \prime }\}_{P} &\rightarrow &A_{1u}\oplus
A_{2u}\oplus E_{u}\oplus E_{u},  \label{eq:symmetry_(s,s,s)}
\end{eqnarray}
where $\{s^{\prime },s^{\prime \prime },s^{\prime \prime \prime }\}_{P}$
denotes a set of distinct permutations of products of functions $\phi
_{n}(q) $ with the indicated symmetry. The characters for every irrep, as
well as some of the basis functions for them, are given in Table~\ref
{tab:D3d}.

It is instructive to carry out a straightforward calculation based on
perturbation theory of first order. For the first energy levels we have
\begin{eqnarray}
E_{1A_{1g}} &=&\frac{3\pi ^{2}}{4}+\frac{2\left( \pi ^{2}-6\right) }{\pi ^{2}%
}\lambda +O(\lambda ^{2})  \nonumber \\
E_{1A_{2u}} &=&\frac{3\pi ^{2}}{2}+\frac{162\pi ^{4}-729\pi ^{2}-4096}{81\pi
^{4}}\lambda +O(\lambda ^{2})  \nonumber \\
E_{1E_{u}} &=&\frac{3\pi ^{2}}{2}+\frac{162\pi ^{4}-729\pi ^{2}+2048}{81\pi
^{4}}\lambda +O(\lambda ^{2})  \nonumber \\
E_{2A_{1g}} &=&\frac{9\pi ^{2}}{4}+\frac{2\left( 81\pi ^{4}-243\pi
^{2}-2048\right) }{81\pi ^{4}}\lambda +O(\lambda ^{2})  \nonumber \\
E_{1E_{g}} &=&\frac{9\pi ^{2}}{4}+\frac{2\left( 81\pi ^{4}-243\pi
^{2}+1024\right) }{81\pi ^{4}}\lambda +O(\lambda ^{2})  \nonumber \\
E_{3A_{1g}} &=&\frac{11\pi ^{2}}{4}+\frac{2\left( 9\pi ^{2}-38\right) }{9\pi
^{2}}\lambda +O(\lambda ^{2})  \nonumber \\
E_{2E_{g}} &=&\frac{11\pi ^{2}}{4}+\frac{2\left( 9\pi ^{2}-38\right) }{9\pi
^{2}}\lambda +O(\lambda ^{2})  \nonumber \\
E_{2A_{2u}} &=&3\pi ^{2}+\frac{2\pi ^{2}-3}{\pi ^{2}}\lambda +O(\lambda ^{2})
\nonumber \\
E_{3A_{2u}} &=&\frac{7\pi ^{2}}{2}+\frac{101250\pi ^{4}-275625\pi
^{2}-2772992}{50625\pi ^{4}}\lambda +O(\lambda ^{2})  \nonumber \\
E_{2E_{u}} &=&\frac{7\pi ^{2}}{2}+\frac{101250\pi ^{4}-275625\pi ^{2}-2048%
\sqrt{466441}}{50625\pi ^{4}}\lambda +O(\lambda ^{2})  \nonumber \\
E_{3E_{u}} &=&\frac{7\pi ^{2}}{2}+\frac{101250\pi ^{4}-275625\pi ^{2}+2048%
\sqrt{466441}}{50625\pi ^{4}}\lambda +O(\lambda ^{2})  \nonumber \\
E_{1A_{1u}} &=&\frac{7\pi ^{2}}{2}+\frac{101250\pi ^{4}-275625\pi
^{2}+2772992}{50625\pi ^{4}}\lambda +O(\lambda ^{2})  \label{eq:E_PT}
\end{eqnarray}
The eigenvalues that are degenerate at $\lambda =0$ exhibit the same parity $%
u$ or $g$. Note that the energy levels $E_{3A_{1g}}$ and $E_{2E_{g}}$
(stemming from the set of quantum numbers $\{3,1,1\}_{P}$) remain degenerate
at first order. We will discuss them in Section~\ref{sec:RR}.

\section{Infinite box}

\label{sec:infinite_box}

The parameter $\lambda $ introduced in Section~\ref{sec:general_trap} is
proportional to the square of an arbitrary length $L$. In the particular
case discussed in Section~\ref{sec:1D_Box} the three particles are confined
into a box of length $2L$ as shown by the boundary conditions (\ref
{eq:box_bc}). Therefore, the infinite-box limit $L\rightarrow \infty $
corresponds to $\lambda \rightarrow \infty $. We appreciate that when $%
\lambda \rightarrow \infty $ the motion of the system center of mass is
unbounded and the eigenfunctions of the Hamiltonian operator are of the form
\begin{equation}
f_{n_{1}\,n_{2}%
\,k}(q_{1},q_{2},q_{3})=g_{n_{1}}(q_{1})g_{n_{2}}(q_{2})e^{ikq_{3}},
\label{eq:eigenfunctions_free_motion}
\end{equation}
where $g_{n}(q)$ is a harmonic-oscillator eigenfunction, $%
n_{1},n_{2}=0,1,\ldots $, and $-\infty <k<\infty $. The eigenvalues behave
asymptotically as
\begin{equation}
\lim\limits_{\lambda \rightarrow \infty }\lambda ^{-1/2}E_{n_{1},n_{2},k}=2%
\sqrt{3}(n_{1}+n_{2}+1).  \label{eq:eigenvalues_free_motion}
\end{equation}
Since the symmetry of the eigenfunction is conserved as $\lambda $ increases
we expect that the small-$\lambda $ eigenfunctions are connected with
\begin{equation}
\left\{
\begin{array}{c}
\cos (kq_{3})g_{n_{1}}(q_{1})g_{n_{2}}(q_{2}) \\
\sin (kq_{3})g_{n_{1}}(q_{1})g_{n_{2}}(q_{2})
\end{array}
\right.  \label{eq:eigenfunctions_free_motion_2}
\end{equation}
instead of (\ref{eq:eigenfunctions_free_motion}). The symmetry of these
functions is also given by equation (\ref{eq:symmetry_(s,s,s)}). For
example, we expect that
\begin{eqnarray}
\lim\limits_{\lambda \rightarrow \infty }\lambda ^{-1/2}E_{1A_{1g}}
&=&\lim\limits_{\lambda \rightarrow \infty }\lambda ^{-1/2}E_{1A_{2u}}=2%
\sqrt{3}  \nonumber \\
\lim\limits_{\lambda \rightarrow \infty }\lambda ^{-1/2}E_{1E_{u}}
&=&\lim\limits_{\lambda \rightarrow \infty }\lambda ^{-1/2}E_{1E_{g}}=4\sqrt{%
3}  \nonumber \\
\lim\limits_{\lambda \rightarrow \infty }\lambda ^{-1/2}E_{1A_{2g}}
&=&\lim\limits_{\lambda \rightarrow \infty }\lambda ^{-1/2}E_{1A_{1u}}=8%
\sqrt{3}.  \label{eq:E_inf_box}
\end{eqnarray}
It is worth noting that $(q_{1},q_{2})$ and $q_{3}$ are bases for the irreps
$E_{u}$ and $A_{2u}$, respectively.

\section{Rayleigh-Ritz variational method with a symmetry-adapted basis set}

\label{sec:RR}

As indicated in earlier papers on the application of group theory to a
variety of problems\cite{F13b,F15a,FG14c,FG14a,FG14b,AF10,AFG14b,AFG14c,F15b}%
, we can obtain approximate eigenvalues and eigenfunctions of $H$ from the
eigenvalues and eigenvectors of the matrix representation $\mathbf{H}^{S}$
of the Hamiltonian operator for every irrep $S$ that we can treat separate
from the other irreps. In the present case the linear combinations of
eigenfunctions of $H_{0}$ (\ref{eq:H0_eigenval_eigenfun}) adapted to the
symmetry species are
\begin{eqnarray}
A_{1g} &:&  \nonumber \\
&&\varphi _{2n-1\,2n-1\,2n-1}  \nonumber \\
&&\frac{1}{\sqrt{3}}\left( \varphi _{2m-1\,2n-1\,2n-1}+\varphi
_{2n-1\,2m-1\,2n-1}+\varphi _{2n-1\,2n-1\,2m-1}\right)  \nonumber \\
&&\frac{1}{\sqrt{6}}\left( \varphi _{2k-1\,2m-1\,2n-1}+\varphi
_{2n-1\,2k-1\,2m-1}+\varphi _{2m-1\,2n-1\,2k-1}\right.  \nonumber \\
&&\left. +\varphi _{2m-1\,2k-1\,2n-1}+\varphi _{2n-1\,2m-1\,2k-1}+\varphi
_{2k-1\,2n-1\,2m-1}\right)  \nonumber \\
&&\frac{1}{\sqrt{3}}\left( \varphi _{2m-1\,2n\,2n}+\varphi
_{2n\,2m-1\,2n}+\varphi _{2n\,2n\,2m-1}\right)  \nonumber \\
&&\frac{1}{\sqrt{6}}\left( \varphi _{2k\,2m\,2n-1}+\varphi
_{2n-1\,2k\,2m}+\varphi _{2m\,2n-1\,2k}\right.  \nonumber \\
&&\left. +\varphi _{2m\,2k\,2n-1}+\varphi _{2n-1\,2m\,2k}+\varphi
_{2k\,2n-1\,2m}\right) ,
\end{eqnarray}

\begin{eqnarray}
A_{2g} &:&  \nonumber \\
&&\frac{1}{\sqrt{6}}\left( \varphi _{2k-1\,2m-1\,2n-1}+\varphi
_{2n-1\,2k-1\,2m-1}+\varphi _{2m-1\,2n-1\,2k-1}\right.  \nonumber \\
&&\left. -\varphi _{2m-1\,2k-1\,2n-1}-\varphi _{2n-1\,2m-1\,2k-1}-\varphi
_{2k-1\,2n-1\,2m-1}\right)  \nonumber \\
&&\frac{1}{\sqrt{6}}\left( \varphi _{2k\,2m\,2n-1}+\varphi
_{2n-1\,2k\,2m}+\varphi _{2m\,2n-1\,2k}\right.  \nonumber \\
&&\left. -\varphi _{2m\,2k\,2n-1}-\varphi _{2n-1\,2m\,2k}-\varphi
_{2k\,2n-1\,2m}\right) ,
\end{eqnarray}

\begin{eqnarray}
E_{g} &:&  \nonumber \\
&&\left\{ \frac{1}{\sqrt{6}}\left( 2\varphi _{2m-1\,2n-1\,2n-1}-\varphi
_{2n-1\,2m-1\,2n-1}-\varphi _{2n-1\,2n-1\,2m-1}\right) ,\right.  \nonumber \\
&&\left. \frac{1}{\sqrt{2}}\left( \varphi _{2n-1\,2m-1\,2n-1}-\varphi
_{2n-1\,2n-1\,2m-1}\right) \right\}  \nonumber \\
&&\left\{ \frac{1}{\sqrt{6}}\left( 2\varphi _{2k-1\,2m-1\,2n-1}-\varphi
_{2n-1\,2k-1\,2m-1}-\varphi _{2m-1\,2n-1\,2k-1}\right) ,\right.  \nonumber \\
&&\left. \frac{1}{\sqrt{2}}\left( \varphi _{2n-1\,2k-1\,2m-1}-\varphi
_{2m-1\,2n-1\,2k-1}\right) \right\}  \nonumber \\
&&\left\{ \frac{1}{\sqrt{6}}\left( 2\varphi _{2m-1\,2k-1\,2n-1}-\varphi
_{2n-1\,2m-1\,2k-1}-\varphi _{2k-1\,2n-1\,2m-1}\right) ,\right.  \nonumber \\
&&\left. \frac{1}{\sqrt{2}}\left( \varphi _{2n-1\,2m-1\,2k-1}-\varphi
_{2k-1\,2n-1\,2m-1}\right) \right\}  \nonumber \\
&&\left\{ \frac{1}{\sqrt{6}}\left( 2\varphi _{2m-1\,2n\,2n}-\varphi
_{2n\,2m-1\,2n}-\varphi _{2n\,2n\,2m-1}\right) ,\frac{1}{\sqrt{2}}\left(
\varphi _{2n\,2m-1\,2n}-\varphi _{2n\,2n\,2m-1}\right) \right\}  \nonumber \\
&&\left\{ \frac{1}{\sqrt{6}}\left( 2\varphi _{2k\,2m\,2n-1}-\varphi
_{2n-1\,2k\,2m}-\varphi _{2m\,2n-1\,2k}\right) ,\left( \varphi
_{2n-1\,2k\,2m}-\varphi _{2m\,2n-1\,2k}\right) \right\}  \nonumber \\
&&\left\{ \frac{1}{\sqrt{6}}\left( 2\varphi _{2m\,2k\,2n-1}-\varphi
_{2n-1\,2m\,2k}-\varphi _{2k\,2n-1\,2m}\right) ,\frac{1}{\sqrt{2}}\left(
\varphi _{2n-1\,2m\,2k}-\varphi _{2k\,2n-1\,2m}\right) \right\} ,  \nonumber
\\
&&
\end{eqnarray}

\begin{eqnarray}
A_{1u} &:&  \nonumber \\
&&\frac{1}{\sqrt{6}}\left( \varphi _{2k\,2m-1\,2n-1}+\varphi
_{2n-1\,2k\,2m-1}+\varphi _{2m-1\,2n-1\,2k}\right.  \nonumber \\
&&\left. -\varphi _{2m-1\,2k\,2n-1}-\varphi _{2n-1\,2m-1\,2k}-\varphi
_{2k\,2n-1\,2m-1}\right)  \nonumber \\
&&\frac{1}{\sqrt{6}}\left( \varphi _{2k\,2m\,2n}+\varphi
_{2n\,2k\,2m}+\varphi _{2m\,2n\,2k}\right.  \nonumber \\
&&\left. -\varphi _{2m\,2k\,2n}-\varphi _{2n\,2m\,2k}-\varphi
_{2k\,2n\,2m}\right) ,
\end{eqnarray}

\begin{eqnarray}
A_{2u} &:&  \nonumber \\
&&\varphi _{2n\,2n\,2n}  \nonumber \\
&&\frac{1}{\sqrt{3}}\left( \varphi _{2m\,2n\,2n}+\varphi
_{2n\,2m\,2n}+\varphi _{2n\,2n\,2m}\right)  \nonumber \\
&&\frac{1}{\sqrt{6}}\left( \varphi _{2k\,2m\,2n}+\varphi
_{2n\,2k\,2m}+\varphi _{2m\,2n\,2k}\right.  \nonumber \\
&&\left. +\varphi _{2m\,2k\,2n}+\varphi _{2n\,2m\,2k}+\varphi
_{2k\,2n\,2m}\right)  \nonumber \\
&&\frac{1}{\sqrt{3}}\left( \varphi _{2m\,2n-1\,2n-1}+\varphi
_{2n-1\,2m\,2n-1}+\varphi _{2n-1\,2n-1\,2m}\right)  \nonumber \\
&&\frac{1}{\sqrt{6}}\left( \varphi _{2k\,2m-1\,2n-1}+\varphi
_{2n-1\,2k\,2m-1}+\varphi _{2m-1\,2n-1\,2k}\right.  \nonumber \\
&&\left. +\varphi _{2m-1\,2k\,2n-1}+\varphi _{2n-1\,2m-1\,2k}+\varphi
_{2k\,2n-1\,2m-1}\right) ,
\end{eqnarray}

\begin{eqnarray}
E_{u} &:&  \nonumber \\
&&\left\{ \frac{1}{\sqrt{6}}\left( 2\varphi _{2m\,2n-1\,2n-1}-\varphi
_{2n-1\,2m\,2n-1}-\varphi _{2n-1\,2n-1\,2m}\right) ,\right.  \nonumber \\
&&\left. \frac{1}{\sqrt{2}}\left( \varphi _{2n-1\,2m\,2n-1}-\varphi
_{2n-1\,2n-1\,2m}\right) \right\}  \nonumber \\
&&\left\{ \frac{1}{\sqrt{6}}\left( 2\varphi _{2k\,2m-1\,2n-1}-\varphi
_{2n-1\,2k\,2m-1}-\varphi _{2m-1\,2n-1\,2k}\right) ,\right.  \nonumber \\
&&\left. \frac{1}{\sqrt{2}}\left( \varphi _{2n-1\,2k\,2m-1}-\varphi
_{2m-1\,2n-1\,2k}\right) \right\}  \nonumber \\
&&\left\{ \frac{1}{\sqrt{6}}\left( 2\varphi _{2m-1\,2k\,2n-1}-\varphi
_{2n-1\,2m-1\,2k}-\varphi _{2k\,2n-1\,2m-1}\right) ,\right.  \nonumber \\
&&\left. \frac{1}{\sqrt{2}}\left( \varphi _{2n-1\,2m-1\,2k}-\varphi
_{2k\,2n-1\,2m-1}\right) \right\}  \nonumber \\
&&\left\{ \frac{1}{\sqrt{6}}\left( 2\varphi _{2m\,2n\,2n}-\varphi
_{2n\,2m\,2n}-\varphi _{2n\,2n\,2m}\right) ,\right.  \nonumber \\
&&\left. \frac{1}{\sqrt{2}}\left( \varphi _{2n\,2m\,2n}-\varphi
_{2n\,2n\,2m}\right) \right\}  \nonumber \\
&&\left\{ \frac{1}{\sqrt{6}}\left( 2\varphi _{2k\,2m\,2n}-\varphi
_{2n\,2k\,2m}-\varphi _{2m\,2n\,2k}\right) ,\right.  \nonumber \\
&&\left. \frac{1}{\sqrt{2}}\left( \varphi _{2n\,2k\,2m}-\varphi
_{2m\,2n\,2k}\right) \right\}  \nonumber \\
&&\left\{ \frac{1}{\sqrt{6}}\left( 2\varphi _{2m\,2k\,2n}-\varphi
_{2n\,2m\,2k}-\varphi _{2k\,2n\,2m}\right) ,\right.  \nonumber \\
&&\left. \frac{1}{\sqrt{2}}\left( \varphi _{2n\,2m\,2k}-\varphi
_{2k\,2n\,2m}\right) \right\} ,
\end{eqnarray}
where $k,m,n=1,2,\ldots $. It is understood that $\varphi _{i\,j\,k}$ means
that the three subscripts are different; equal subscripts are indicated
explicitly as in $\varphi _{i\,i\,i}$ or $\varphi _{i\,j\,j}$.

\begin{figure}[tbp]
\begin{center}
\includegraphics[width=6cm]{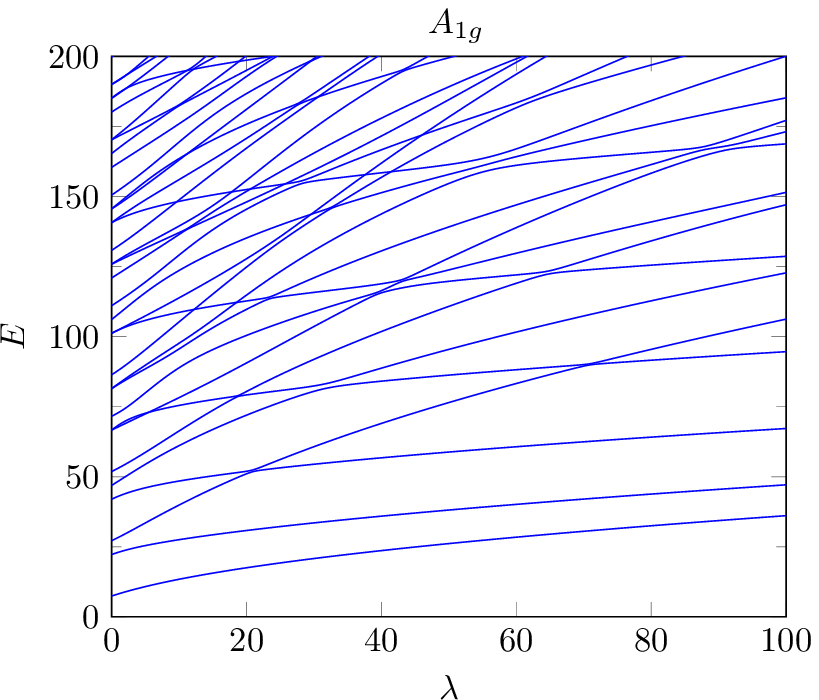} %
\includegraphics[width=6cm]{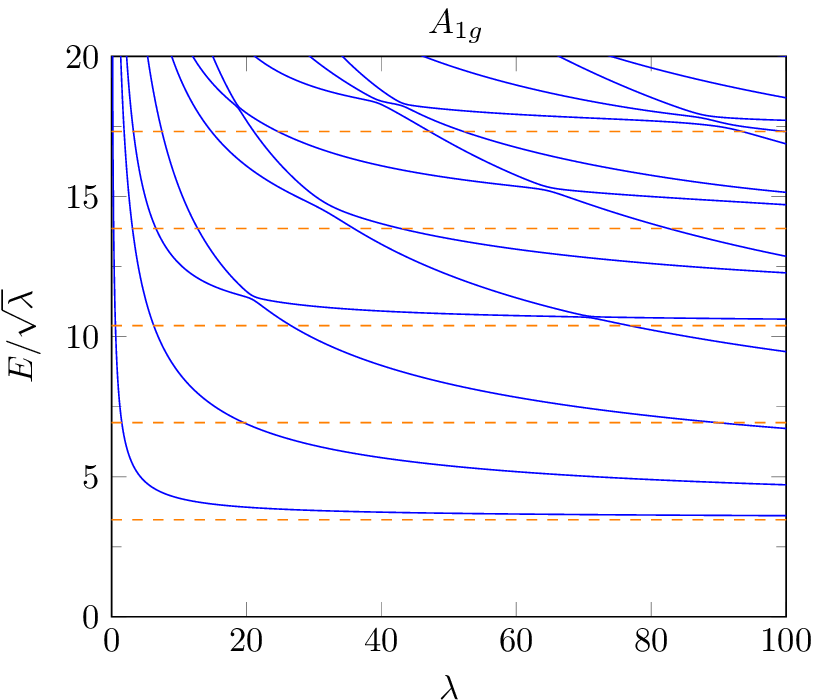}
\par
\end{center}
\caption{Eigenvalues for the irrep $A_{1g}$}
\label{fig:EA1g}
\end{figure}

\begin{figure}[tbp]
\begin{center}
\includegraphics[width=6cm]{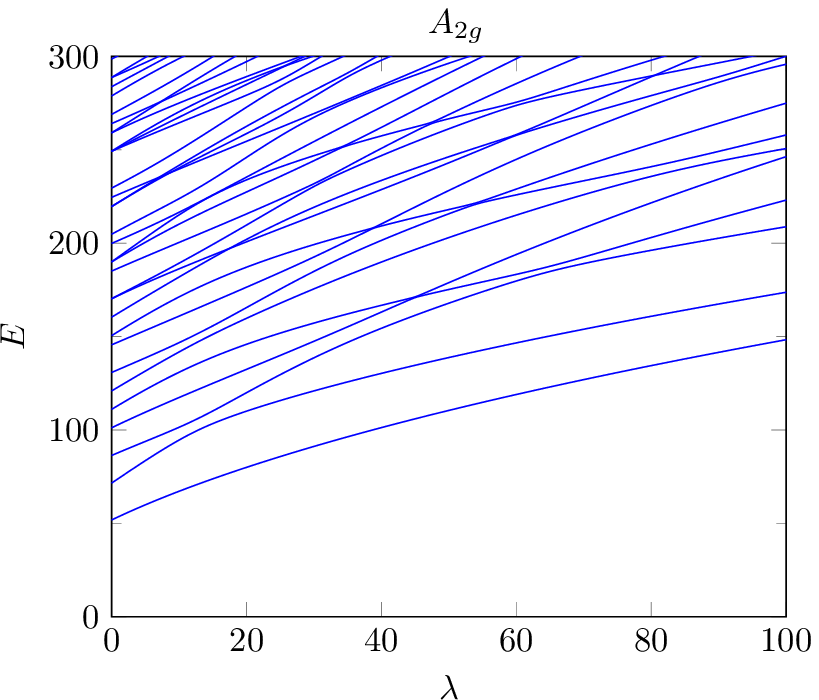} %
\includegraphics[width=6cm]{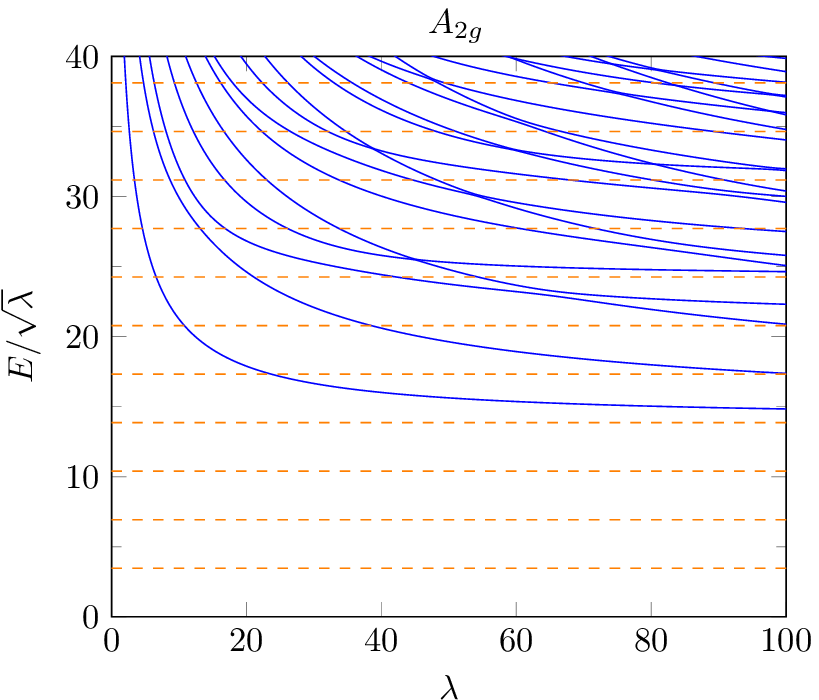}
\par
\end{center}
\caption{Eigenvalues for the irrep $A_{2g}$}
\label{fig:EA2g}
\end{figure}

\begin{figure}[tbp]
\begin{center}
\includegraphics[width=6cm]{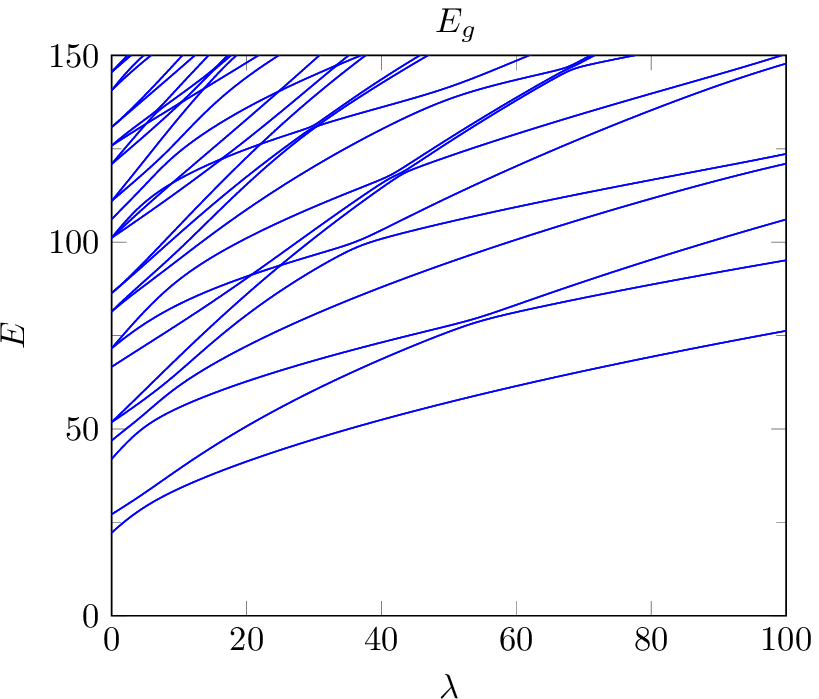} %
\includegraphics[width=6cm]{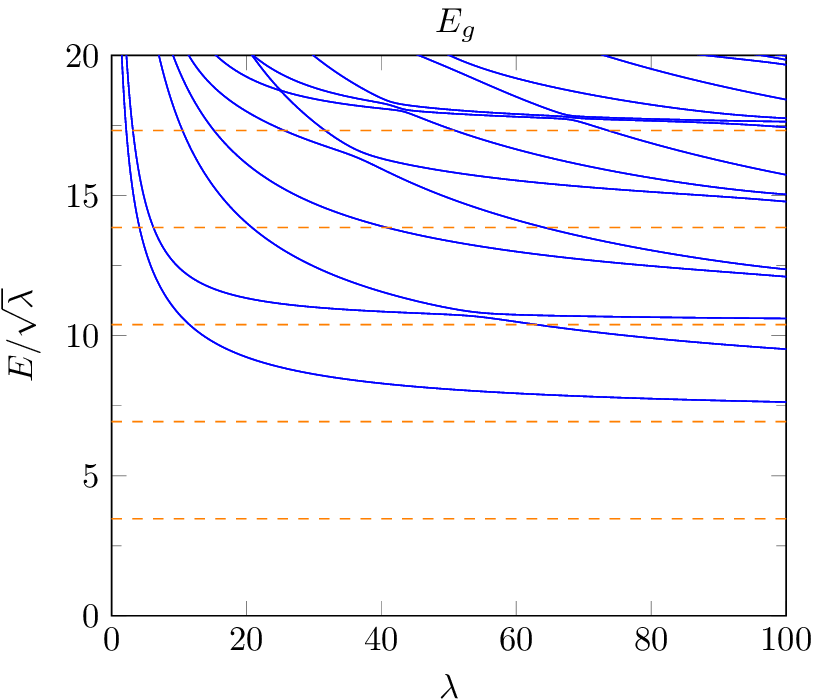}
\par
\end{center}
\caption{Eigenvalues for the irrep $E_{g}$}
\label{fig:EEg}
\end{figure}

\begin{figure}[tbp]
\begin{center}
\includegraphics[width=6cm]{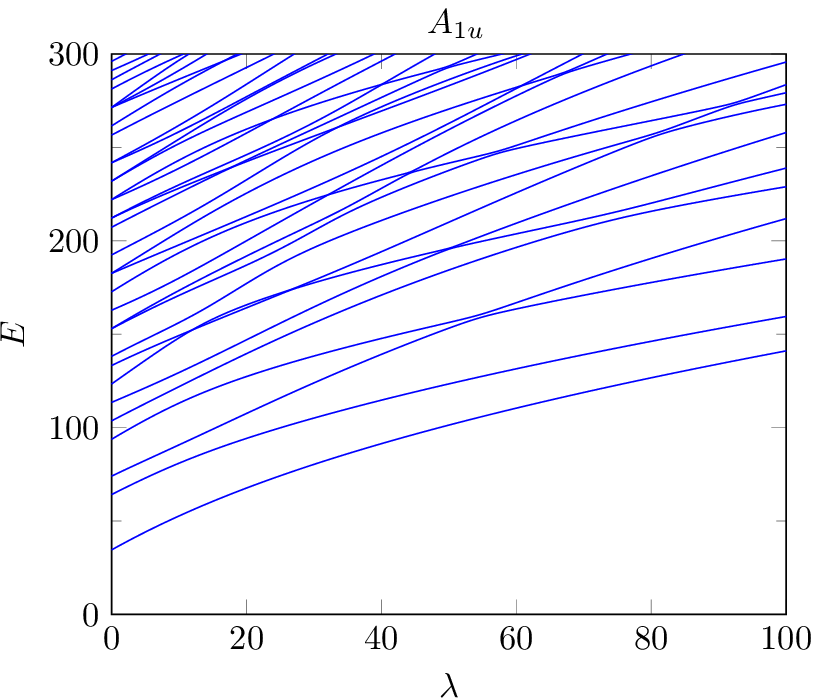} %
\includegraphics[width=6cm]{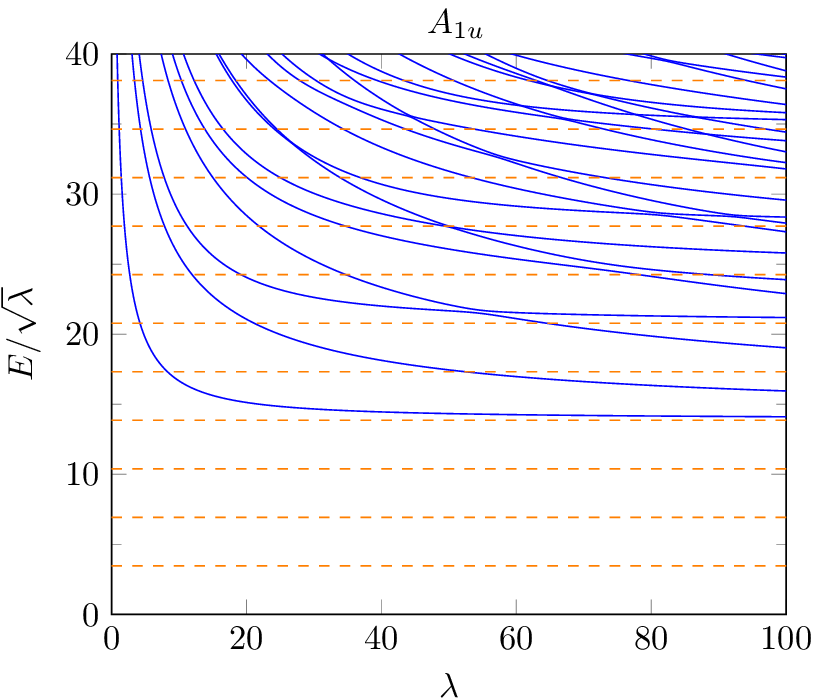}
\par
\end{center}
\caption{Eigenvalues for the irrep $A_{1u}$}
\label{fig:EA1u}
\end{figure}

\begin{figure}[tbp]
\begin{center}
\includegraphics[width=6cm]{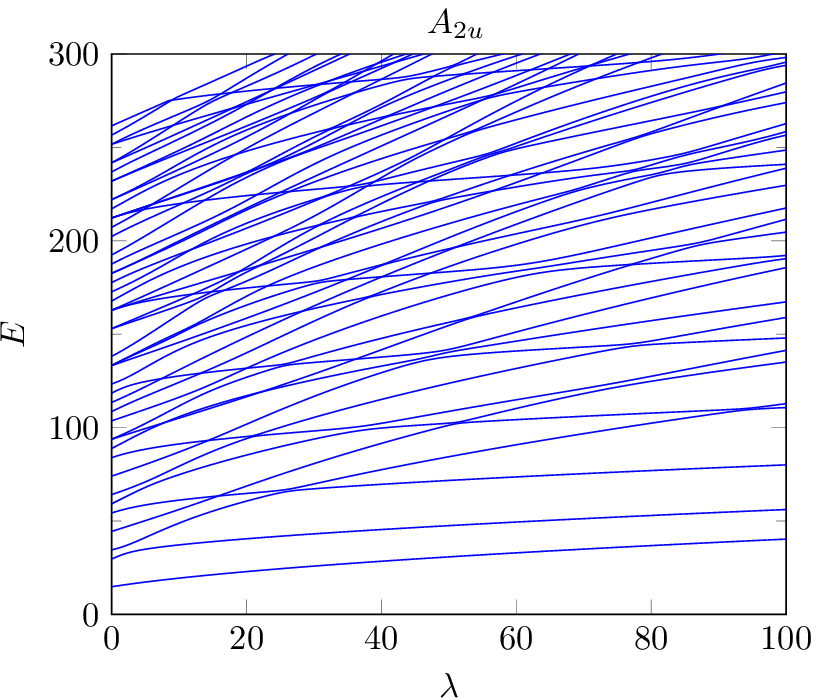} %
\includegraphics[width=6cm]{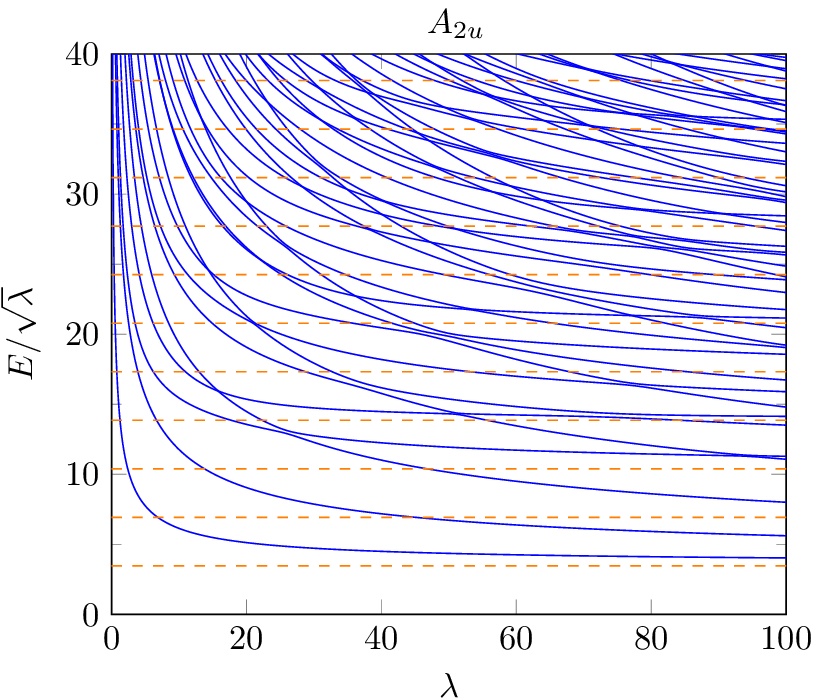}
\par
\end{center}
\caption{Eigenvalues for the irrep $A_{2u}$}
\label{fig:EA2u}
\end{figure}

\begin{figure}[tbp]
\begin{center}
\includegraphics[width=6cm]{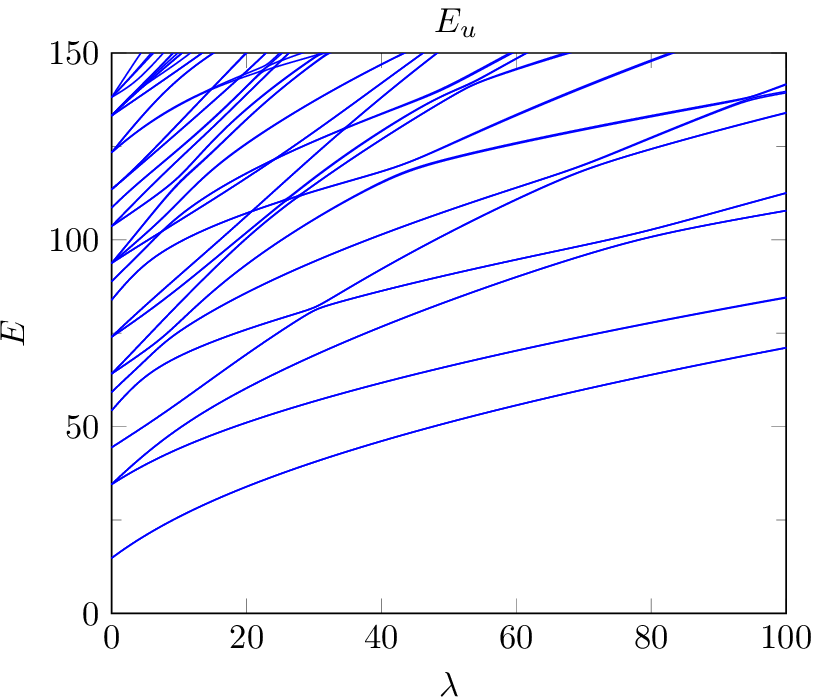} %
\includegraphics[width=6cm]{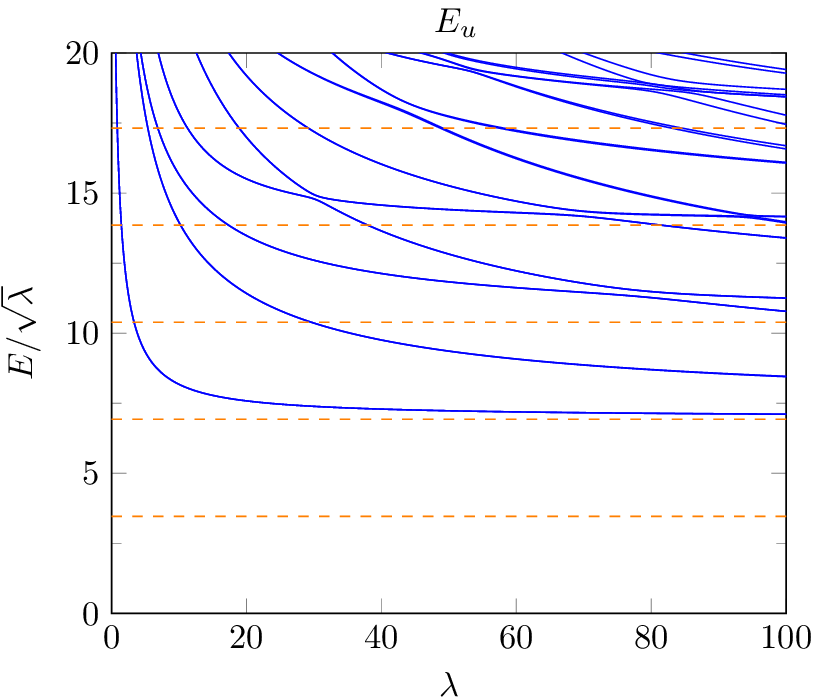}
\par
\end{center}
\caption{Eigenvalues for the irrep $E_{U}$}
\label{fig:EEu}
\end{figure}

In this paper we have carried out a diagonalization of the matrix
representation of the Hamiltonian operator with $1000$ basis functions of
each symmetry species. Figs.~\ref{fig:EA1g}-\ref{fig:EEu} show $E(\lambda )$
and $\lambda ^{-1/2}E(\lambda )$ for the first eigenvalues of each irrep.
The latter illustrate the conclusions drawn in Section~\ref{sec:infinite_box}
and in particular equations (\ref{eq:E_inf_box}). To facilitate the analysis
we have drawn horizontal dashed lines that mark the limits (\ref
{eq:eigenvalues_free_motion}).

\begin{figure}[tbp]
\begin{center}
\includegraphics[width=6cm]{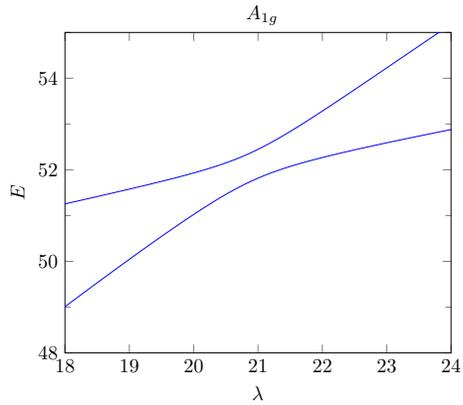}
\par
\end{center}
\caption{Third and fourth eigenvalues of symmetry $A_{1g}$}
\label{fig:EA1g_c}
\end{figure}

It is well known that eigenvalues of the same symmetry do not cross but
exhibit what is commonly known as avoided crossings\cite{F14}. Figs.~\ref
{fig:EA1g}-\ref{fig:EEu} exhibit some clear avoided crossings and others
that appear to be actual crossings because of the scale of the figures. As
an example in Fig.~\ref{fig:EA1g_c} we show the third and fourth states of
symmetry $A_{1g}$ on a finer scale to make it clear that they in fact
undergo an avoided crossing.

\begin{figure}[tbp]
\begin{center}
\includegraphics[width=6cm]{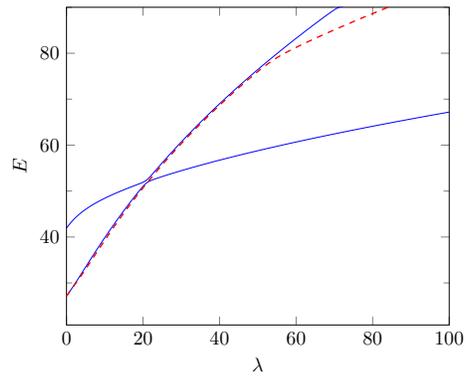}
\par
\end{center}
\caption{Third and fourth eigenvalues of symmetry $A_{1g}$ (blue, solid
line) and second pair of eigenvalues of symmetry $E_g$ (red, dashed line) }
\label{fig:3A1g-2Eg}
\end{figure}

Fig.~\ref{fig:3A1g-2Eg} shows a most interesting feature of the spectrum of
this system of identical particles. The eigenvalues $E_{2E_{g}}$ and $%
E_{3A_{1g}}$ are almost degenerate for small $\lambda $ (in Section~\ref
{sec:RR} we have seen that they share the same energy of order zero and
first order correction). However, the slope of the energy level $E_{3A_{1g}}$
changes dramatically at the avoided crossing $\lambda _{c}$ with the level $%
E_{4A_{1g}}$ but the slope of $E_{2E_{g}}$ does not change and the latter
level approaches and remains close to $E_{4A_{1g}}$ for $\lambda >\lambda
_{c}$.

\section{Conclusions}

\label{sec:conclusions}

It is interesting that a simple one-dimensional model exhibits such a rich
symmetry. We have shown that group theory is remarkably useful for the
interpretation of the results produced by approximate methods like
perturbation theory and the variational method. The treatment of each irrep
separately from the others renders the analysis of the spectrum considerably
simpler. For example, the avoided crossings between levels of equal symmetry
can be studied separate from the crossings between levels of different
symmetry. The interplay among the energy levels $E_{3A_{1g}}$, $E_{4A_{1g}}$
and $E_{2E_{g}}$ is most interesting and one wonders if such behaviour also
takes place in more realistic models of quantum-mechanical problems.

Since the symmetry of the problem does not change with $\lambda $ we can
resort to group theory to predict the connection of the states in the small-$%
\lambda $ and large-$\lambda $ regimes embodied in Eq.~\ref{eq:E_inf_box}.
These results were confirmed by the accurate Rayleigh-Ritz variational
calculation in Section~\ref{sec:RR}.

Group theory is also useful for the calculation of matrix elements of
observables because we already know beforehand which of them vanish. This is
the reason why we can treat each irrep separate from the others in the
Rayleigh-Ritz calculation. The prediction of zero matrix elements is also
the basis, for example, of the well known selection rules in spectroscopic
transitions\cite{H62,C90}.

In closing we mention that the combination of group theory and perturbation
theory has recently led us to the conclusion that the well known parity-time
symmetry is less easily broken that other forms of the more general
space-time symmetry\cite{AFG14b}. This result is of great relevance for the
study of non-Hermitian Hamiltonians.

\section*{Acknowledgements}

All the figures in this paper have been produced by means of the Tikz package%
\cite{TikZ}.

F.M.F. would like to thank Dr. N. L. Harshman for enlightening discussions
on group theory and systems of identical particles.

In order to facilitate the application of group theory to systems of
identical particles F.M.F. resorted to some of the programs contributed (for
other purposes) by the Derive User Group\cite{DUG}. In particular, F.M.F.
would like to thank Joseph B\"{o}hm for useful discussions about such
programs.

International Derive User Group, http://www.austromath.at/dug/.

\begin{table}[tbp]
\caption{Character table for $D_{3d}$ point group}
\label{tab:D3d}
\begin{tabular}{l|rrrrrr|l|l}
& $E$ & $2C_3$ & $3C_2$ & $i$ & $2S_6$ & $3\sigma_d$ &  &  \\
$A_{1g}$ & 1 & 1 & 1 & 1 & 1 & 1 &  & $x^2+y^2+z^2$ \\
&  &  &  &  &  &  &  & $xy+yz+zx$ \\
$A_{2g}$ & 1 & 1 & -1 & 1 & 1 & -1 &  &  \\
$E_{g}$ & 2 & -1 & 0 & 2 & -1 & 0 &  & $(2z^2-x^2-y^2,x^2-y^2)$, \\
&  &  &  &  &  &  &  & $(2yz-xy-xz,xy-xz)$ \\
$A_{1u}$ & 1 & 1 & 1 & -1 & -1 & -1 &  &  \\
$A_{2u}$ & 1 & 1 & -1 & -1 & -1 & 1 & $x+y+z$ &  \\
$E_{u}$ & 2 & -1 & 0 & -2 & 1 & 0 & $(2z-x-y,x-y)$ &
\end{tabular}
\end{table}

\end{document}